\documentclass[letterpaper,pra,floatfix,twocolumn,aps,10pt]{revtex4-1}

\usepackage{
	amssymb, 
	amsthm, 
	graphicx, 
	xspace,
	overpic,
	ifthen,
	verbatim, 
	subfigure, 
	appendix,
	ifthen,
	nicefrac, 
	color, 
	tensor, 
	etoolbox, 
	centernot, 
	MnSymbol, 
	amsmath,
	bm
} 

\newcommand{\ket}[1]{\lvert #1 \rangle  }

\newcommand{\braket}[2]{  \langle #1 \vert #2 \rangle  }

\newcommand{\projector}[1]{  \lvert #1 \rangle \langle #1 \rvert }

\newcommand{\abs}[1]{| #1 |} 
\newcommand{\Abs}[1]{\left| #1 \right|}

\newcommand{\avg}[1]{\langle #1 \rangle} 
\newcommand{\Avg}[1]{\left\langle #1 \right\rangle}

\newcommand{\dd}{\mathrm{d}} 
\newcommand{\ddd}[1]{\! \mathrm{d}#1 \,} 

\newcommand{\refstyle}{2}  
\newcommand{\eqrefstyle}{2}  
\newcommand{\Jeqref}[1] {\ifnum\refstyle=1{Eq.~(\ref{#1})}\else{equation (\ref{#1})}\fi}
\newcommand{\Jeqsref}[1] {\ifnum\refstyle=1{Eqs.~(\ref{#1})}\else{equations (\ref{#1})}\fi}
\newcommand{\EQref}[1] {\ifnum\eqrefstyle=1{(\ref{#1})}\else{\Jeqref{#1}}\fi}
\newcommand{\EQsref}[1] {\ifnum\eqrefstyle=1{(\ref{#1})}\else{\Jeqsref{#1}}\fi}
\newcommand{\Eqref}[1]  {\ifnum\eqrefstyle=3{\Jeqref{#1}}\else{(\ref{#1})}\fi}
\newcommand{\Eqsref}[1]  {\ifnum\eqrefstyle=3{\Jeqsref{#1}}\else{(\ref{#1})}\fi}
\newcommand{\Figref}[1]{\ifnum\refstyle=1{Fig.~\ref{#1}}\else{figure~\ref{#1}}\fi}
\newcommand{\Figsref}[1]{\ifnum\refstyle=1{Figs.~\ref{#1}}\else{figures~\ref{#1}}\fi}
\newcommand{\Appref}[1]{\ifnum\refstyle=1{App.~\ref{#1}}\else{appendix~\ref{#1}}\fi}
\newcommand{\Appsref}[1]{\ifnum\refstyle=1{Apps.~\ref{#1}}\else{appendices~\ref{#1}}\fi}
\newcommand{\Chref}[1]{\ifnum\refstyle=1{Ch.~\ref{#1}}\else{chapter~\ref{#1}}\fi}
\newcommand{\Chsref}[1]{\ifnum\refstyle=1{Chs.~\ref{#1}}\else{chapters~\ref{#1}}\fi}
\newcommand{\Secref}[1]{\ifnum\refstyle=1{Sec.~\ref{#1}}\else{section~\ref{#1}}\fi}
\newcommand{\Secsref}[1]{\ifnum\refstyle=1{Secs.~\ref{#1}}\else{sections~\ref{#1}}\fi}
\newcommand{\Refref}[1]{\ifnum\refstyle=1{Ref.~\cite{#1}}\else{reference~\cite{#1}}\fi}
\newcommand{\Refsref}[1]{\ifnum\refstyle=1{Refs.~\cite{#1}}\else{references~\cite{#1}}\fi}

\newcommand{\draftmode}{1}    
\newcommand{\notetoself}[1]{\ifnum \draftmode=1 {\color[rgb]{0,0,0.8} [#1]} \fi}  
\newcommand{\cuttext}[1]{\ifnum \draftmode=1 {\color[rgb]{0,0.5,0} [#1]} \fi}  
\newcommand{\warntext}[1]{{\ifnum \draftmode=1 \color[rgb]{0.9,0.6,0} \else  \color{black} \fi} #1}  
\usepackage{phonetic}
\usepackage[english]{babel}
\pdfoutput=1

\renewcommand{\draftmode}{0} 
\newcommand{\Nor}{\ensuremath{\mathcal{N}}} 
\newcommand{\Nk}{\ensuremath{N}} 
\newcommand{\Eth}{\ensuremath{\mathcal{D}}}
\newcommand{\mdm}{\ensuremath{m_\mathrm{DM}}}  

\newcommand{\dm}{DM} 

\begin{document}


%
\title{Direct detection of classically undetectable dark matter through quantum decoherence}
\date{\today}
\author{C.~Jess~Riedel}\
\affiliation{IBM Watson Research Center, Yorktown Heights, NY, USA}


\begin{abstract}
Although various pieces of indirect evidence about the nature of dark matter have been collected, its direct detection has eluded experimental searches despite extensive effort.  If the mass of dark matter is below 1 MeV, it is essentially imperceptible to conventional detection methods because negligible energy is transferred to nuclei during collisions.  Here I propose directly detecting dark matter through the quantum decoherence it causes rather than its classical effects such as recoil or ionization.  I show that quantum spatial superpositions are sensitive to low-mass dark matter that is inaccessible to classical techniques.  This provides new independent motivation for matter interferometry with large masses, especially on spaceborne platforms.  The apparent dark matter wind we experience as the Sun travels through the Milky Way ensures interferometers and related devices are directional detectors, and so are able to provide unmistakable evidence that decoherence has galactic origins.
\end{abstract}

\maketitle

\section{Introduction}

It has been almost 80 years since dark matter (DM) was first proposed to explain the observed orbital velocities within galaxies and galaxy clusters.  Much additional evidence for its existence has accumulated in the subsequent decades, but it has always been indirect and essentially gravitational.  Existing experiments, observations, and theoretical preferences form a complicated thicket of conditional restrictions on any potential theory of \dm{} \cite{Bartelmann2010, Beringer2012}, but model-independent constraints are still rare and very valuable.  Most desirable would be the \emph{direct} detection of \dm{}, that is, local experiments here on Earth that observe the interaction of the \dm{} particle with the well-known electrons, protons, and neutrons that compose the normal matter we see around us.  

Based just on the movement of luminous matter in the galaxy and the virial theorem, we infer that \dm{} forms a halo which has a density of roughly $\rho \sim 0.4$ GeV/cm$^3$ \cite{Catena2010} in the neighborhood of the solar system.  The \dm{} particles should follow a non-relativistic Maxwellian velocity distribution centered around $v_0 \sim 230$ km/s in the galactic rest frame with a cutoff at the galactic escape velocity $v_\mathrm{e} \sim 600$ km/s \cite{Lewin1996}.  Given this, any hypothetical mass $\mdm$ for the \dm{} particle then fixes its rough interaction rate with a particle of normal matter (either an electron or a nucleon) as a function of a scattering cross-section $\sigma$: $\Gamma_0 \sim v_0 \sigma \rho / \mdm$.  I will concentrate on spin-independent elastic scattering with nucleons, which has been extensively studied in the context of direct detection. 

Conventional \dm{} direct detection experiments consist essentially of a large container of normal matter (e.g.\ liquid xenon) which is carefully watched for the tiny effects of an elastic collision with the \dm{}  particle such as recoil, vibration, heating, or ionization.  These techniques rely on there being sufficient energy transfer from the \dm{} to the target that the state of the target is substantially changed.

For a collision with a target particle of mass $M$, the energy deposited is no more than $2 \mdm^2 v_0^2/M$.  Most often the target is an atomic nucleus, for which direct detection experiments are sensitive down to a few keV of energy. This corresponds to a sensitivity to \dm{} masses greater than a few GeV and lines up well with the Lee-Weinberg bound \cite{Lee1997}, which constrains the most popular form of weakly interacting massive particle (WIMP) dark matter to a mass of at least 2 GeV.

However, the most natural WIMP models have been challenged by galactic N-body simulations and negative direct experimental searches.  It is prudent to allow for more general possibilities, and there are many proposals for sub-GeV mass such as WIMPless \cite{Feng2008b}, ``MeV'' \cite{Hooper2008}, bosonic super-WIMP \cite{Pospelov2008}, or asymmetric \cite{Falkowski2011} dark matter.
Calorimetry experiments which look for bulk heating rather than individual collisions have been able to explore down to 10 MeV in a modest cross-section range \cite{Erickcek2007}.  Scattering off of electrons---if it occurs---can probe masses as low as 1 MeV \cite{Essig2012a, Graham2012} because the lighter electrons absorb more energy and also have a lower detectable energy threshold than nuclei. But conventional direct detection techniques probably cannot do any better than this, at least without making specific, model-dependent assumptions. A 1 MeV \dm{} candidate will deposit about an eV when colliding with an electron and less than $10^{-3}$ eV when colliding with a nucleus.  For a keV candidate, the values are $10^{-6}$ and $10^{-9}$ eV, respectively. 

In a classical universe, sub-MeV dark matter would be ghostly.  Its dynamics could be strongly influenced by normal matter, but it would leave little trace.  More precisely, such \dm{} would be undetectable using classical measurement strategies \cite{BraginskyText, Giovannetti2004} in that phase-space localized states of normal matter would not be appreciably perturbed by collisions.

In this article, I propose searching for low-mass \dm{} by observing the quantum decoherence \cite{Zurek2003, SchlosshauerText} it causes rather than its direct influence on normal matter.  This technique is notable because it is sensitive to \dm{} masses that are generally considered to be undetectable.  To my knowledge it is the first proposal for using decoherence in this manner.

\begin{figure} [tb!]
    \centering 
  \newcommand{\pbwidthfactor}{0.95}
    \includegraphics[width=\pbwidthfactor\columnwidth]{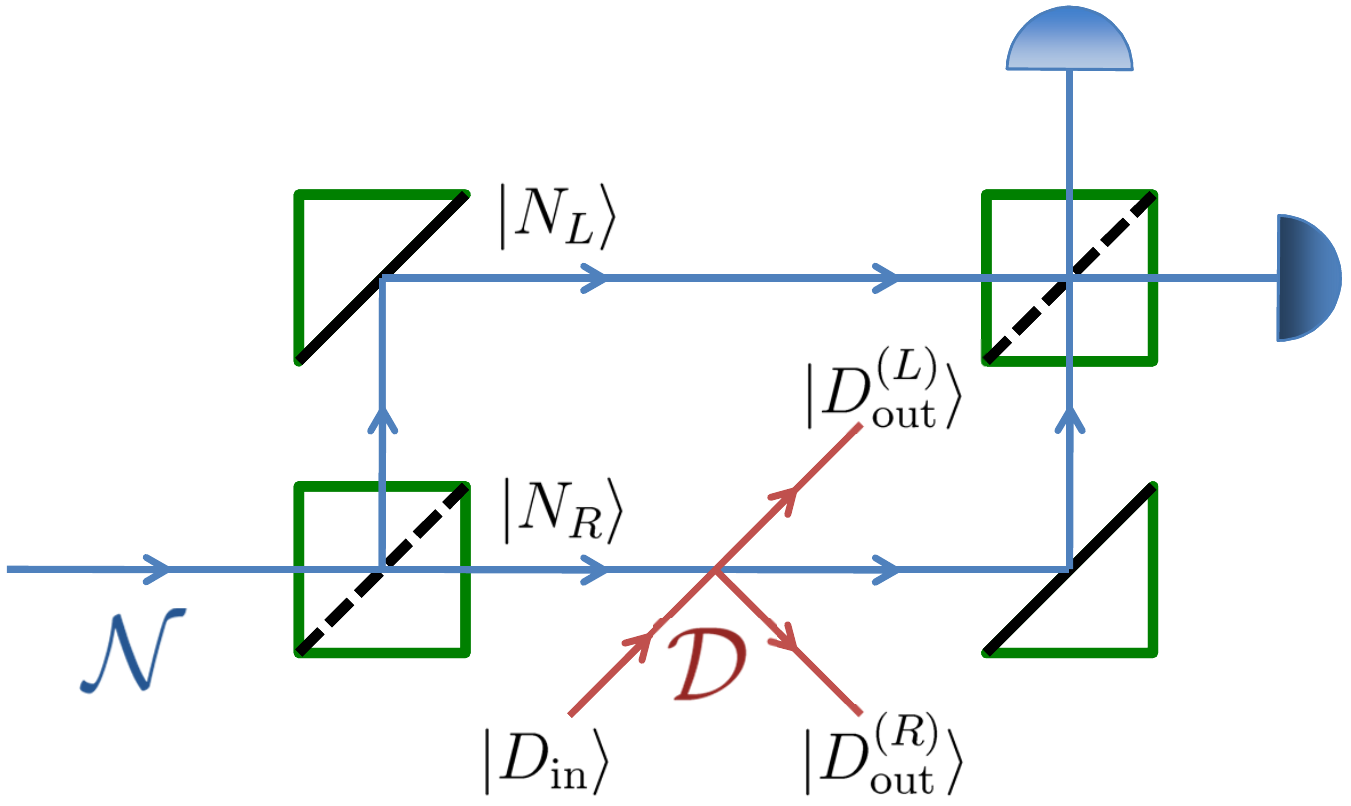}
  \caption{\textbf{Decoherence detection with a Mach-Zehnder interferometer.} The target $\Nor$ is placed in a coherent superposition of spatially displaced wavepackets which each travel a separate path and then are recombined.  In the absence of the dark matter $\Eth$, the interferometer is tuned so that $\Nor$ will be detected at the bright port with near unit probability, and at the dim port with near vanishing probability.  However, if the dark matter $\Eth$ scatters off of $\Nor$, these two paths will decohere and $\Nor$ will be detected at the dim port 50\% of the time.}
  \label{mz_diagram}
\end{figure}

\section{Collisional decoherence by dark matter}

As an alternative to conventional direct detection methods, consider a Mach-Zehnder atom interferometer that takes advantage of the de-Broglie-wave nature of matter, \Figref{mz_diagram}.  An atom $\Nor$ is prepared in a coherent superposition $\ket{\Nk_{L}} + \ket{\Nk_{R}}$ of two wavepackets, one taking the left path and one taking the right path, with something functioning as a beam splitter.  These wavepackets are allowed to propagate over some length, and then they are recombined with a second splitter.  Assuming the spread of the wavepackets is negligible and the splitters are properly aligned, the sensors effectively measure $\Nor$ in the basis $\{\ket{\Nk_\pm} =  \ket{\Nk_{L}} \pm \ket{\Nk_{R}} \}$.  The atom ends up at one ``bright'' port, corresponding to the measurement outcome $\ket{\Nk_+}$, with near unit probability, and at the other ``dim'' sensor, corresponding to $\ket{\Nk_-}$, with near vanishing probability.

Now we allow for the possibility of \dm{} interacting with the atom while it is in interferometer. If we let the state $\ket{D_\varnothing}$ represent the absence of \dm{}, then the evolution is trivial when \dm{} is not present,
\begin{align}
\Big[\ket{\Nk_{L}} + \ket{\Nk_{R}} \Big] \ket{D_\varnothing} \to \Big[\ket{\Nk_{L}} + \ket{\Nk_{R}} \Big] \ket{D_\varnothing} ,
\end{align}
so measuring in the basis $\{\ket{\Nk_\pm} \}$ gives outcome $\ket{\Nk_+}$ with certainty, as before.  But suppose the \dm{} particle approaches in some state $\ket{D_\mathrm{in}}$ and decoheres the superposition by scattering off the atom,
\begin{align}
\Big[\ket{\Nk_{L}} + \ket{\Nk_{R}} \Big] \ket{D_\mathrm{in}} \to \ket{\Nk_{L}}\ket{D^{(L)}_\mathrm{out}} + \ket{\Nk_{R}} \ket{D^{(R)}_\mathrm{out}} ,
\end{align}
into the conditional states $\ket{D^{(L)}_\mathrm{out}}$ and $\ket{D^{(R)}_\mathrm{out}}$ with $\braket{D^{(L)}_\mathrm{out}}{D^{(R)}_\mathrm{out}} \approx 0$, thereby recording which-path information.  When $\mdm \ll 1$ GeV, the phase-space localized wavepackets $\ket{\Nk_{L}}$ and $\ket{\Nk_{R}}$ of the atom are not significantly perturbed following the scattering event.  But a measurement in the basis $\{\ket{\Nk_\pm} \}$ now gives equal probability of either outcome.  When the dim sensor clicks (which it will do half the time), this gives direct evidence of the existence of $\Eth$ \emph{even when the dark matter transfers negligible momentum to $\Nor$}.

(Of course, decoherence is ubiquitous \cite{SchlosshauerText}. Convincingly identifying dark matter as the source of decoherence and eliminating alternative explanations is discussed in section \ref{sec:anom-decoh}.)

\begin{figure} [tb!]
  \centering 
\newcommand{\nerdfactor}{0.95}
\includegraphics[width=\nerdfactor\columnwidth]{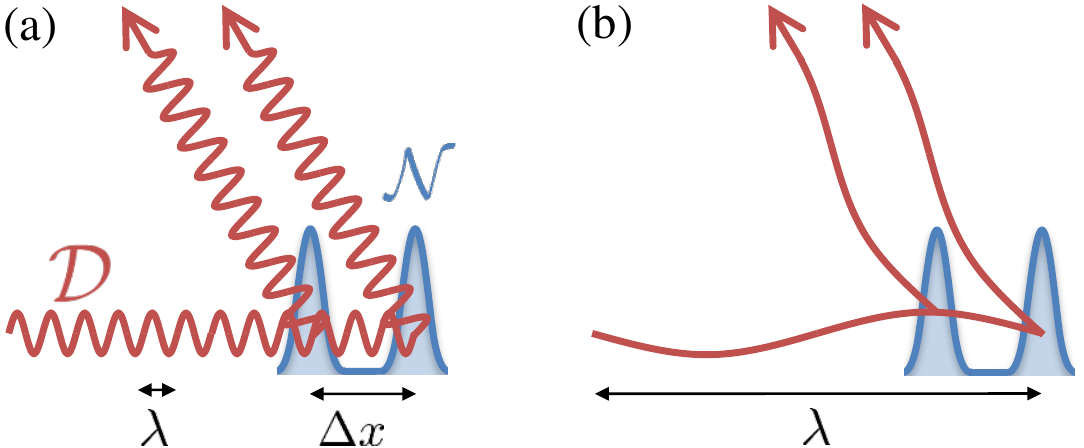}
\caption{\textbf{Decoherence by dark matter with different de Broglie wavelengths.} An atom $\Nor$ in a superposition of spatial extent $\Delta x$ is decohered by a \dm{} particle $\Eth$ of wavelength $\lambda$.  For s-wave (hard-sphere) scattering, $\lambda$ is also the wavelength associated with the typical momentum transfer to the \dm{} particle.  (a) In the short-wavelength limit $\lambda \ll \Delta x$, a single scattering event completely decoheres: $\gamma = \braket{D^{(L)}_\mathrm{out}}{D^{(R)}_\mathrm{out}} \approx 0$.  (b) For longer wavelengths, the \dm{} cannot easily ``see'' the superposition and it takes many scattering events to decohere \cite{Joos1985}: $\abs{\braket{D^{(L)}_\mathrm{out}}{D^{(R)}_\mathrm{out}}} = 1 - \epsilon$, with $\epsilon$ small, but $\abs{\gamma} = \abs{\braket{D^{(L)}_\mathrm{out}}{D^{(R)}_\mathrm{out}}}^N  \approx e^{-N \epsilon} \approx 0$ for sufficiently large $N$.}
  \label{fig:wave}
\end{figure}

The ability of a single \dm{} particle to decohere an atom through elastic scattering is determined by the overlap $\braket{D^{(L)}_\mathrm{out}}{D^{(R)}_\mathrm{out}}$ of the conditional dark-matter post-scattering states.    This overlap, in turn, is strongly affected by the typical de Broglie wavelength $\lambda_0 = 2 \pi \hbar / v_0 \mdm$ of the \dm{} particle compared to the spatial separation $\vec{\Delta x}$ between the two wavepackets.  See \Figref{fig:wave}.
So long as the occupation number of the \dm{} is much less than unity, it can be treated as a fixed number of identically distributed but distinguishable particles.  The state of the atom $\Nor$  in the $\{\ket{\Nk_L},  \ket{\Nk_R} \}$ basis after a time $T$ is
\begin{align}
\rho_\Nor = \frac{1}{2}\begin{pmatrix} 1 & \gamma \\ \gamma^* & 1 \end{pmatrix}
\end{align}
where $\gamma = \exp [ - \int_0^T \!\! \dd t \,  F(\Delta \vec{x})]$ is the decoherence factor and $F(\vec{\Delta x})$ is given by \cite{ SchlosshauerText}
\begin{align}\begin{split}
\label{decoh-rate}
F(\vec{\Delta x}) =& \int \dd \vec{q} \, n(\vec{q}) \frac{q}{\mdm} \int \dd  \hat{r} \\
& \times \left\{1 - \exp[i (\vec{q} -  q\hat{r}) \cdot \vec{\Delta x}/\hbar]\right\} \abs{f( \vec{q}, q \hat{r})}^2.
\end{split}\end{align}
Above, $\vec{q}$ is the incoming \dm{} momentum, $n(\vec{q})$ is the spatially homogeneous distribution function (phase space number density) of \dm{} and $\abs{f(\vec{q}_\mathrm{in}, \vec{q}_\mathrm{out})}^2 = \dd \sigma / \dd \Omega$ is the differential cross-section.  This is collisional decoherence, and it was first analyzed in detail by Joos and Zeh \cite{Joos1985}.  (See references \cite{Gallis1990, Diosi1995, Hornberger2003a, Hornberger2006, Adler2006} for extensions and corrections, and  references \cite{SchlosshauerText, Adler2006} for a discussion of the historical development of \Eqref{decoh-rate}.)

The atomic superposition is fully decohered when $\abs{\gamma} \ll 1$, that is $\mathrm{Re} \, F(\Delta x) \gtrsim 1/T$.  The density matrix is then diagonal, and both possible outcomes of the measurement are equiprobable.  For general $q$, the angular integrals over $\hat{q}$ and $\hat{r}$ in \Eqref{decoh-rate} can only be done by assuming a form for the differential cross-section. The scattering should be effectively elastic for the nucleon because the \dm{} is far too feeble to excite internal nuclear states.  Furthermore, the s-wave component of the partial-wave expansion is expected to dominate because of the very long de Broglie wavelength of sub-MeV \dm{} \cite{SquiresText}. So long as the cross-section does not vary too quickly with momentum $q$, it is reasonable to take $\dd \sigma / \dd \Omega = \sigma/4\pi$ to be constant.

\section{Massive superpositions and the coherent scattering enhancement}

The toy detector in \Figref{mz_diagram} only works if the flux of \dm{} is high enough such that at least one \dm{} particle will usually scatter off of an atom wavepacket while it is in the interferometer.
Because of the rarity of collisions, interferometry with single atoms has little hope of being sensitive to \dm{}.  There are at least two ways to increase the likelihood of a scattering event, and hence increase the possibility of detection.  First, the time over which the superposition is maintained can be increased by lengthening the interferometer arms or slowing down the atom. Each unit of time contributes an independent opportunity for a scattering event.  Second, and more powerfully, one can superpose ever larger clusters of atoms.  That is, construct a \emph{matter} interferometer with targets $\Nor$ which are as large as possible.  As the number of nucleons composing $\Nor$ increases, each contributes an independent decoherence factor.  This multiplies the effective decoherence rate $F_\mathrm{R} \equiv \mathrm{Re} \,F(\vec{\Delta x})$ by the total number of nucleons.

Moreover, for \dm{} with sufficiently long de Broglie wavelength (i.e.\ low $\mdm$), there is a significant enhancement to the total spin-independent scattering cross-section through \emph{coherent elastic scattering} with nucleons, a process that is well known from small-angle scattering of neutrons and x-rays \cite{SquiresText} and investigations into the possibility of detecting relic neutrinos \cite{Smith1991}.  In this process, multiple nucleons can contribute coherently to the amplitude (rather than to the probability) of the same \dm{} out state because the outgoing \dm{} does not ``know'' which nucleon it has scattered from.  The nucleons recoil together uniformly, as in the M\"ossbauer effect.  The enhancement is maximum when the entire target is smaller than the reduced de Broglie wavelength $\lambdabar \equiv \lambda/2 \pi$ of the \dm{} and the total scattering rate is then proportional to the \emph{square} of the number of nucleons in the target.  At the other extreme, when the \dm{} wavelength is much smaller than the nuclear scale, there is no enhancement compared to normal incoherent scattering (for which the total cross-section is linear in nucleon number).  In the intermediate regime, the enhancement is roughly proportional to the number of nucleons which fit in the \emph{coherent scattering volume}, a sphere of diameter $\lambdabar$ (figure \ref{fig:coherscatt}). See the Appendix for a complete explanation and further discussion.

To achieve interference of large objects with ever smaller de Broglie wavelengths, modern time-domain interferometers can require a time interval proportional to the size of the object superposed  \cite{Nimmrichter2011a, Haslinger2013}.  When combined with the coherent scattering enhancement, the \dm{} sensitivity can scale like the \emph{cube} of the quoted mass of the superposed object. Although this is partially a testament to the difficulty of superposing large objects, it also means that investing in larger masses yields big dividends.  Happily, recent progress in the size of superposed objects in matter interferometry has been stunning, with clear fringe patterns produced when interfering molecules composed of up to 430 atoms and in excess of 6,000 amu \cite{Gerlich2011}.  Future prospects are even stronger \cite{Nimmrichter2011a, Romero-Isart2011, Kaltenbaek2012, kaltenbaek2013testing, Haslinger2013, asenbaum2013cavity}, and these have great potential for discovery. Techniques already being deployed are expected to achieve superpositions exceeding $10^6$ amu \cite{Nimmrichter2011a, Haslinger2013}.

\section{Unitary phase shift from dark matter wind}

Unlike earlier examples of collisional decoherence \cite{Joos1985, Gallis1990, SchlosshauerText}, the scattering \dm{} environment is not distributed isotropically because of the significant speed with which the sun orbits the galactic center.  The apparent \dm{} ``wind'' drives the imaginary part of $F(\vec{\Delta x})$ to a non-zero value.  

$F_\mathrm{I} \equiv \mathrm{Im} \,F(\vec{\Delta x})$ changes the phase of the decoherence factor $\gamma$.  When $F_\mathrm{R} = \mathrm{Re} \,F(\vec{\Delta x})$ is large compared to $T$, this doesn't matter; $\gamma$ vanishes regardless.  However, when $F_\mathrm{R}$ is small, $\abs{\gamma} \approx 1$ and the density matrix is then given by $\rho \approx \projector{\psi}$ for $\ket{\psi} = \ket{\Nk_L} + \gamma \ket{\Nk_R}$.  The state has not been decohered and is still pure. Instead, the \dm{} environment has acted unitarily on the normal matter by applying a position-dependent phase. A position-dependent phase per unit time is simply a coherent \emph{classical} force, and it is natural that this net force is only non-zero when the \dm{} momentum distribution is anisotropic.  Note that this phase shift is still the result of complete \dm{} scattering events, and is distinct from a possible index of refraction arising from forward scattering.  (The former is second-order in the interaction strength, while the latter is first-order but receives no enhancement from coherent scattering.)  

Although the \dm{} wind does not spatially displace the wavepackets $\ket{\Nk_L}$ and $\ket{\Nk_R}$ by measurable amounts, it is known that such forces can still be detected using interferometry.  In this sense the force of the wind is analogous to the force of gravity in the famous neutron interferometry experiments of Colella et al.\ \cite{Colella1975}. This is a quantum-enhanced measurement \cite{BraginskyText, Giovannetti2004} and, although it has not been used to detect new particles, it is the basis behind many existing weak-force experiments (e.g.\ Refs.\ \cite{Peters1999, Dimopoulos2008}).  In such experiments, the semiclassical approximation applies wherein the classical force is modeled as a unitary influence on the quantum state of the test masses.  No entanglement between the test mass and the force mediators is possible.  In our case, the details of the \dm{} and the interferometer determine the ratio $F_\mathrm{I} / F_\mathrm{R}$, and this governs the transition from the well-known coherent case (which can be modeled as a unitary evolution of the target) to the decoherent case introduced here (which cannot).

\begin{figure} [tb!]
  \centering 
\newcommand{\nerdfactor}{0.6}
\includegraphics[width=\nerdfactor\columnwidth]{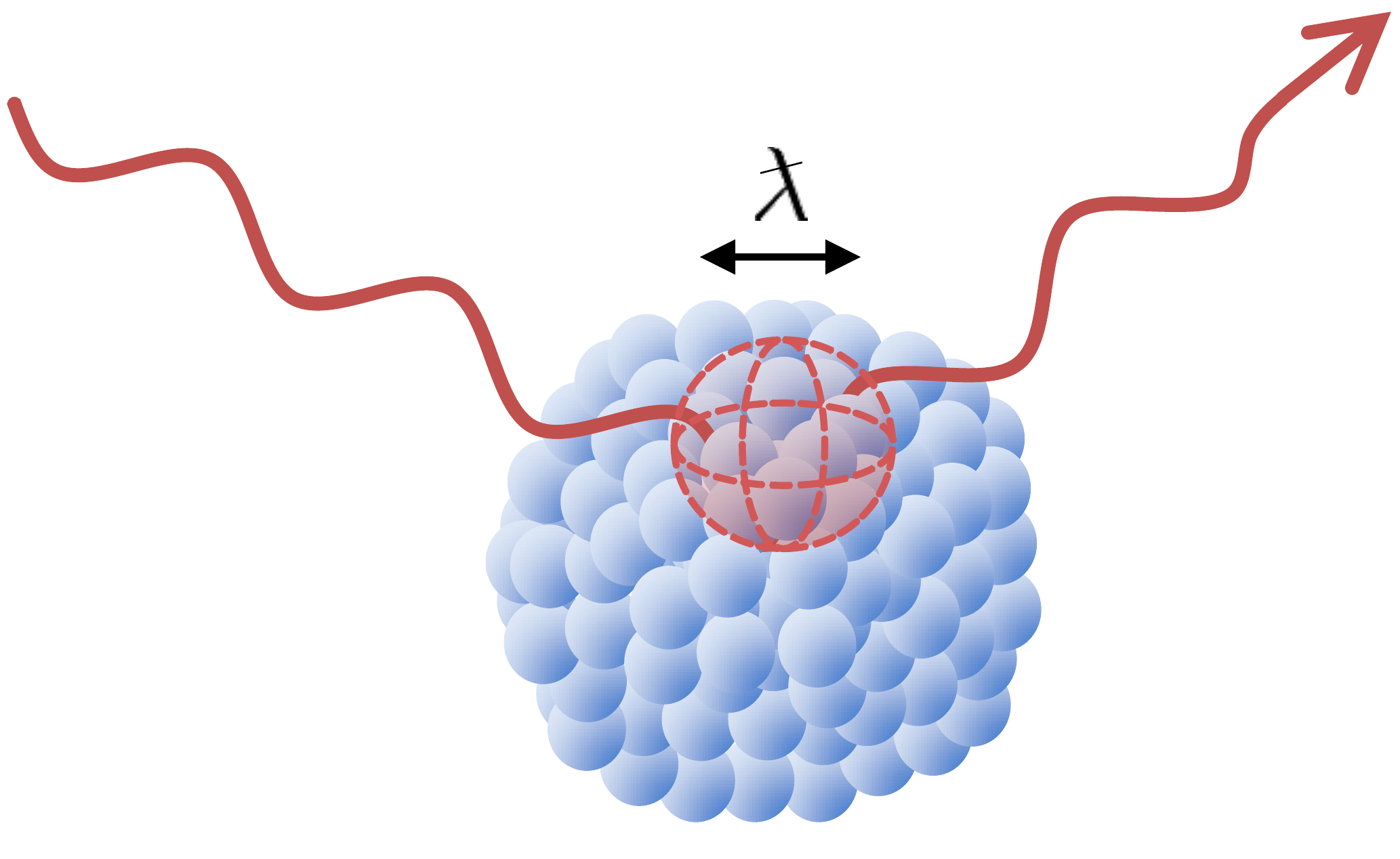}
\caption{\textbf{Coherent scattering over multiple nuclei.} For sufficiently small $\mdm$, the \dm{} wavelength is too long to resolve the individual nuclei in a cluster of atoms. Instead, the \dm{} scatters coherently from multiple nuclei which recoil together uniformly.  The multiplicative enhancement to the total cross-section due to this effect (compared to normal incoherent scattering) is well-approximated by the number of nucleons that fit inside the coherent scattering volume, a sphere of diameter $\lambdabar = \lambda/2 \pi$. If the \dm{} de Broglie wavelength is sufficiently long, the volume can contain the entire target so that the total cross-section is proportional to the target mass squared. See the Appendix for details.}
  \label{fig:coherscatt}
\end{figure} 

\begin{figure*} [tb!]
  \centering 
\newcommand{\mathematicascale}{0.51}
\includegraphics[scale=\mathematicascale]{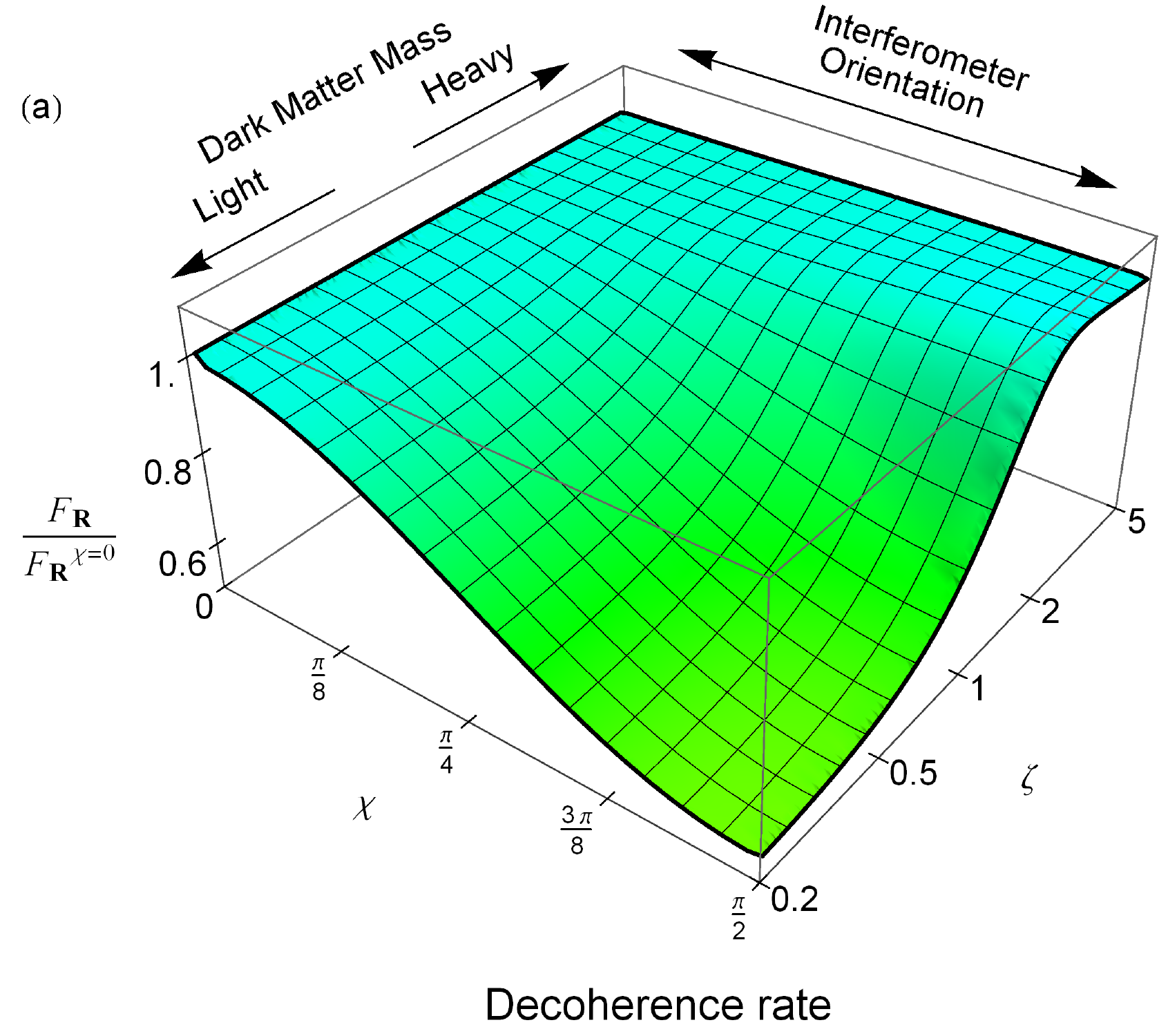}
\includegraphics[scale=\mathematicascale]{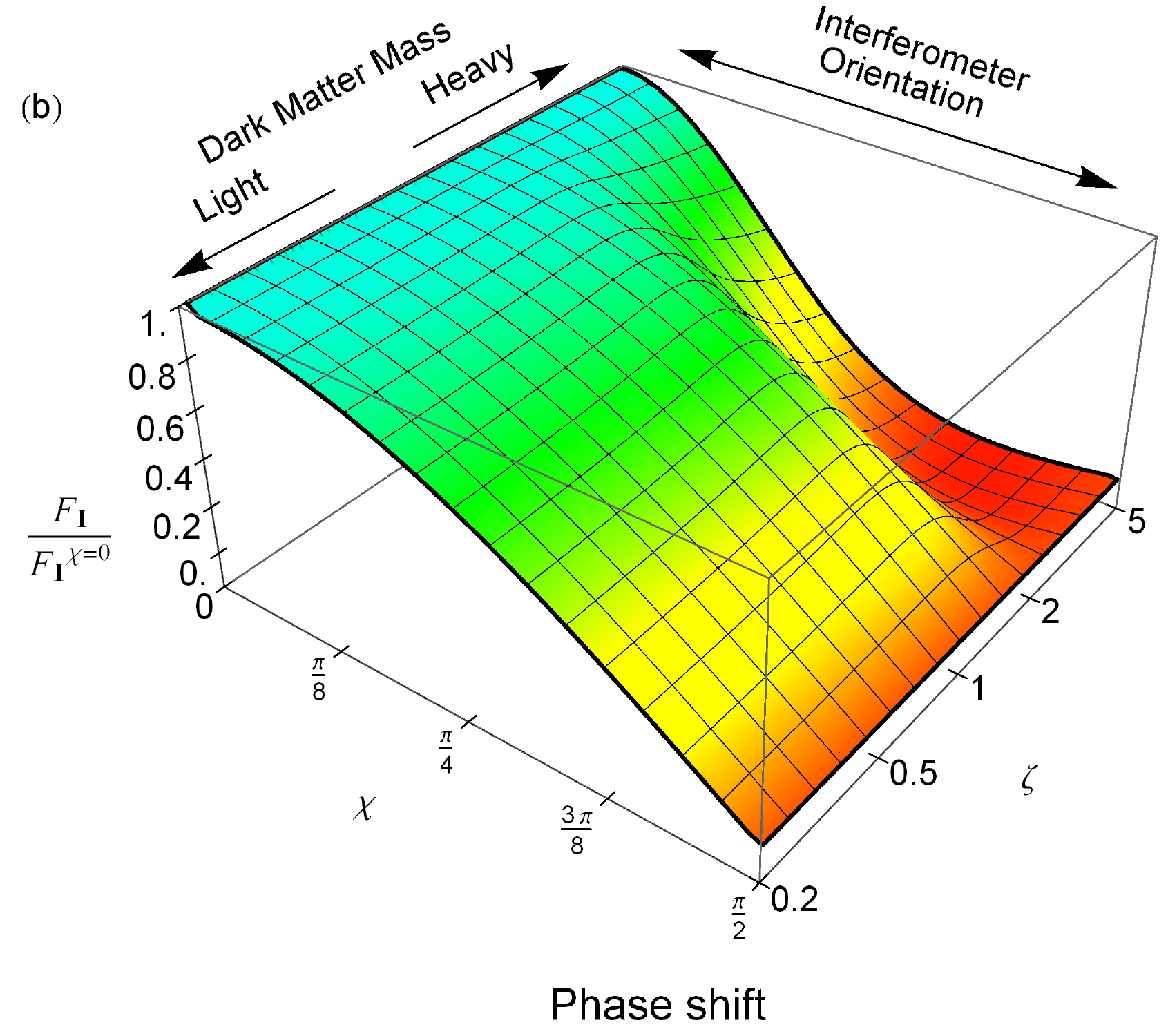}
  \caption{\textbf{Visibility of the dark matter wind.}  The \dm{} momentum distribution seen by an experiment is concentrated in a direction opposite the velocity $\vec{v}_{\mathrm{Earth}}$ of the Earth in the galactic rest frame.	The sensitivity of the superposition to interactions with \dm{} is determined by the angle $\chi$ between $\vec{v}_{\mathrm{Earth}}$ and the spatial displacement $\vec{\Delta x}$ between the two parts of the superposition.  $\chi$ can be adjusted by changing $\vec{\Delta x}$, i.e.\ rotating the interferometer. (a) The real part of $F$ controls the decoherence of the superposition. It is plotted as a function of $\chi$ and $\zeta = \mdm v_0 \Delta x / \hbar = \Delta x / \lambdabar_0$. (It is normalized to its value at $\chi = 0$.) This is how the decoherence strength would fluctuate as an interferometer is rotated with respect to the \dm{} wind.  Values for $\chi > \pi/2$ are given by $\mathrm{Re} \, F_{\pi - \chi} = \mathrm{Re} \, F_{\chi}$.  For \dm{} wavelengths much shorter than the size of the superposition ($\zeta \to \infty$), there is no dependence on orientation because a single collision event causes complete decoherence.   (b) The imaginary part of $F$ controls the coherent phase shift between the arms of the interferometer due to the weak force applied by the wind. Values for larger $\chi$ are given by $F_\mathrm{I} (\pi - \chi) = - F_\mathrm{I}(\chi)$.  Although the phase shift has $\chi$ dependence for all $\zeta$, decoherence always prevents the observation of this shift in the short-wavelength limit: $F_\mathrm{I}(\chi) / F_\mathrm{R}(\chi) \to 0$ as $\zeta \to \infty$.
}
  \label{fig:chi}
\end{figure*} 

Since interferometers cannot measure a constant phase shift between their two arms, the force must vary to be observable.  There are therefore two related motivations for modulating the \dm{} flux: (1) if anomalous decoherence is detected, its functional dependence on parameters which control the hypothetical \dm{} scattering gives evidence that the decoherence is in fact due to \dm{}, and (2) if the \dm{} wind applies only a coherent phase shift, rather than decoherently dephasing, then some variation in time is necessary to observe this shift at all.  Modulation techniques are discussed in the next section.

\section{Implications of anomalous decoherence}
\label{sec:anom-decoh}

Uncontrolled decoherence from many sources is often the primary barrier to constructing experiments which establish grossly non-classical states.  When an experimentalist succeeds in suppressing decoherence from one source (e.g. phonons), a new weaker source (e.g. blackbody radiation) is often revealed which must then be dealt with in turn.  This staircase of decoherence must be descended until \emph{all} sources of decoherence have been driven below some level.  Sometimes every source of decoherence can be understood through careful theoretical analysis, but not all experiments are so amenable.  \emph{Anomalous decoherence} is decoherence that resists theoretical understanding, and it by itself is certainly not convincing evidence for new physics.  Confidently attributing anomalous decoherence to \dm{}, and furthermore extracting the physical parameters of \dm{}, will require more care.

It's worth emphasizing first that the inverse situation is not ambiguous; the experimental verification of a superposition immediately excludes dark matter parameter space.  This is because different sources of decoherence will contribute independently and additively to the decoherence rate (i.e. they will contribute their own multiplicative decoherence factors).  A superposition can only survive if all sources of decoherence have been eliminated.

But if the experiment shows signs of decoherence despite all known conventional sources being eliminated, a next step is to adjust the experimental parameters and see whether the resulting degree of decoherence agrees with predictions based on the hypothesized \dm{} source.  Widening or lengthening the arms of the interferometer, or adjusting the speed of the target, should change the interference fringe visibility through the parameters $T$ and $\vec{\Delta x}$.  Changing the isotopic composition of the targets would change the \dm{} cross-section of the nuclei without affecting other (extra-nuclear) sources of decoherence.  The most convincing evidence for establishing the source of decoherence will come by manipulating the source itself, i.e.\ by modulating the \dm{} flux. This can be done in several complimentary ways.

First, predictable natural variations in the apparent flux, such as those due to the Earth's orbital motion around the sun, may be exploited.  Such a signal can be confounded by other effects with a similar period, but this technique nevertheless has an extensive history \cite{Michelson1887, Bernabei2003, Aalsethetal2011}.  

Second, the incoming \dm{} may be directly shielded from reaching the detector.  This can be done using normal materials, such as lead or concrete, for almost all of the parameter space we will consider.  For $\sigma = 10^{-29}\mathrm{cm}^2$, the attenuation length $\ell_\mathrm{Pb}$ in lead is about a meter.  (With regard to shielding, there will be a coherent scattering enhancement over the nucleus but not the bulk; see the Appendix. Also note that shields could have complicated effects on the DM flux, like thermalization within the normal matter.)   Although the Earth's crust is not a particularly efficient shield owing to the smaller average atomic mass, shielding for $\sigma \gtrsim 10^{-31}\mathrm{cm}^2$ could still be accomplished by operating the interferometer in an underground laboratory at depths $\sim 2000$ m below the surface. For even lower cross-sections, the entire Earth can be used as a giant \dm{} ``windscreen'' so long as $\sigma \gtrsim 10^{-35} \mathrm{cm}^2$. Anomalous decoherence with a 24-hour period could be investigated by moving the experiment elsewhere on the Earth's surface and looking for the appropriate shift in the time-dependence of the signal.

Third, outside of the short wavelength limit (i.e.\ when it takes multiple scattering events to fully decohere), the orientation of the superposition with respect to the direction of the \dm{} wind gives an order-unity modulation of the effect.  Decoherence is usually maximized when the wind is parallel to the separation vector $\vec{\Delta x}$.  Alternatively, when the evolution of the target is roughly coherent, the phase shift due to the wind flips sign as the orientation is rotated. See \Figref{fig:chi}. This means interferometers are naturally \emph{directional} \dm{} detectors, which are known to be highly desirable \cite{Ahlenetal2010} in part because they provide unmistakable evidence that a signal is of galactic origin.  The Earth's daily rotation guarantees that this directional variation will be visible even to fixed terrestrial experiments.

\begin{figure*} [tb!]
  \centering 
\newcommand{\weirdfactor}{1.07}
\includegraphics[height=\weirdfactor\columnwidth]{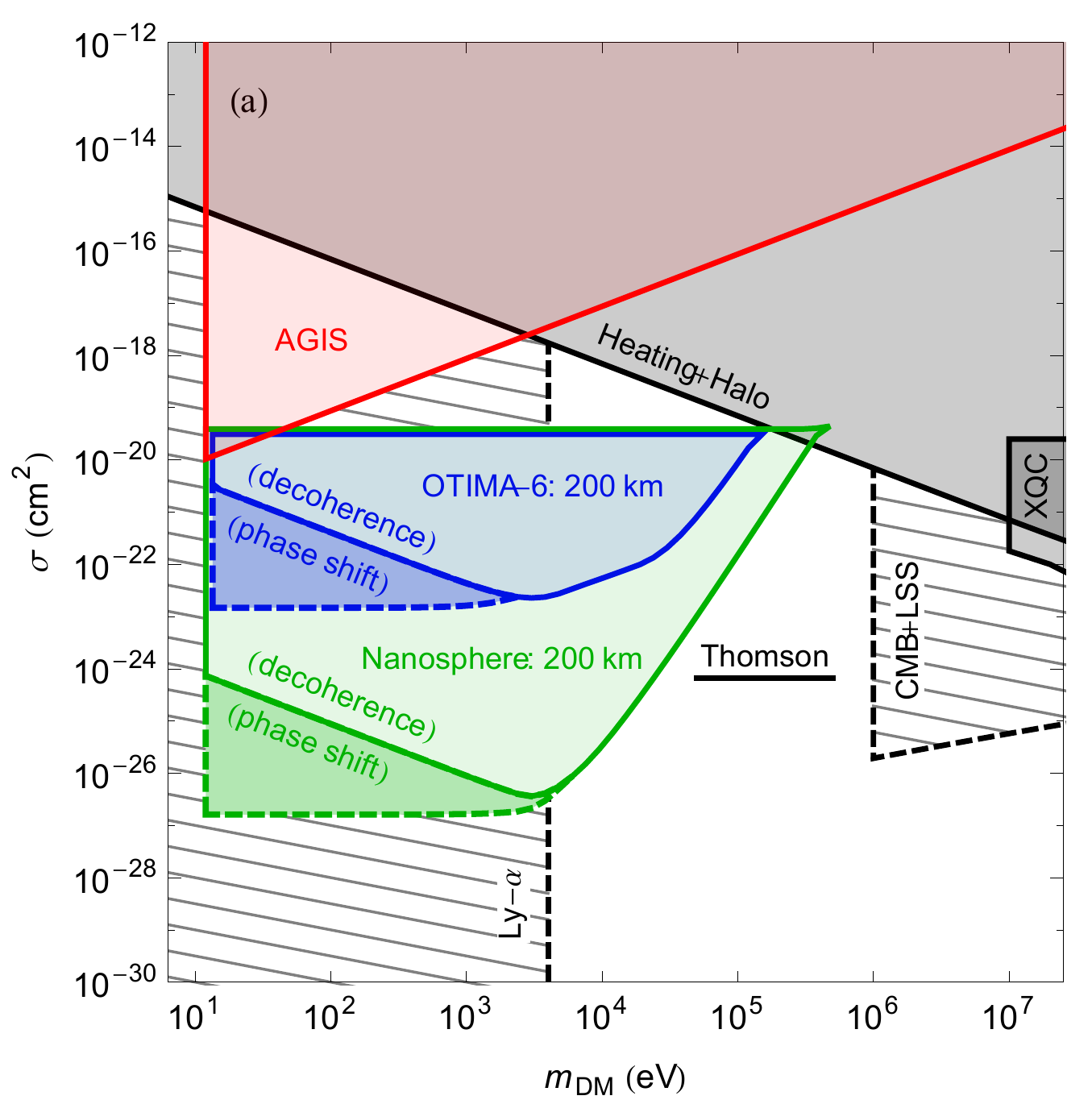}
\includegraphics[height=\weirdfactor\columnwidth]{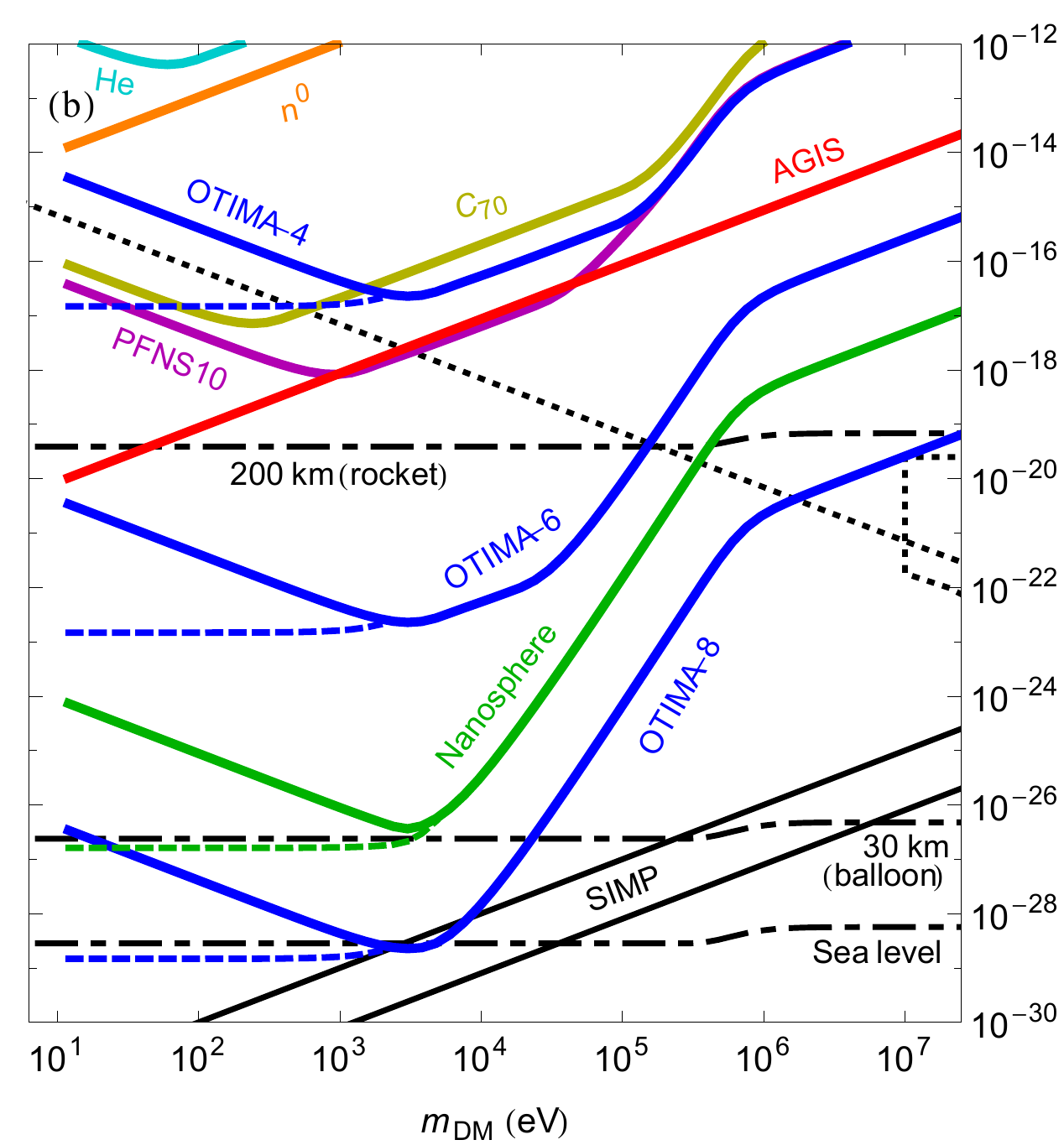}
\caption{\textbf{The sensitivity of several existing and proposed superposition experiments to the spin-independent nucleon scattering cross-section of dark matter, compared with existing constraints.}  (a) Gray shaded regions are robustly excluded by the X-ray Quantum Calorimetry experiment (``XQC'' \cite{Erickcek2007}) and heating and halo stability arguments (``Heating+Halo'' \cite{Chivukula1990}).  Hatched regions are incompatible with thermal \dm{} models due to observations of the cosmic microwave background with large scale structure data (``CMB+LSS'' \cite{Chen2002}) and the Lyman-$\alpha$ forest (``Ly-$\alpha$'' \cite{Viel2008}).   Solid colored lines bound regions where \dm{} would cause decoherence in three proposed experiments: a satellite-based atom interferometer (``AGIS'' \cite{hogan2011atomic}), a 40 nm diameter optically-trapped silicon nanosphere (``Nanosphere'' \cite{Romero-Isart2011}), and the OTIMA interferometer with clusters of gold of mass $10^6$ amu (``OTIMA-$6$'' \cite{Nimmrichter2011a}).  A successful AGIS satellite would set new exclusion limits on \dm{} where its sensitivity dips below the heating and halo stability bound for $\mdm \lesssim 3$ keV.  On the other hand, the OTIMA and nanosphere experiments would be shielded from \dm{} by the atmosphere if operated at sea level, so exclusion regions illustrate the sensitivity at an altitude of 200 km. 
The darker regions bordered by colored dashed lines indicates where the coherent phase shift due to the \dm{} wind could be observed without being overwhelmed by decoherence. For comparison, the Thomson cross-section of the electron (``Thomson'') is $6.65 \times 10^{-25}\, \mathrm{cm}^2$. (b)  On top of the existing exclusions (now black dotted lines), the colored lines give the lower limits on the sensitivities of existing interferometers with helium atoms (``He'' \cite{Carnal1991}), cold neutrons (``$\mathrm{n}^0$'' \cite{Zeilinger1988}), fullerenes (``$\mathrm{C_{70}}$'' \cite{Brezger2002}), and the large organic molecule $\mathrm{C_{60}[C_{12}F_{25}]_{10}}$ (``PFNS10'' \cite{Gerlich2011}).  Also shown are sensitivities for the AGIS satellite, the nanosphere experiment, and the OTIMA interferometer with three choices of gold cluster mass. (``OTIMA-$N$'' denotes cluster mass $10^N$ amu for $N=4,6,8$, although the last is not feasible for an Earth-bound experiment.)  The border is defined by an $e$-fold suppression of the interference fringes:  $\abs{\gamma} = 1/e$.  Sensitivity increases dramatically for larger target masses.  When an experiment is operated within the Earth's atmosphere, there is a potential to detect \dm{} only where the sensitivity dips below the dashed-dotted line corresponding to the degree of shielding at the relevant altitudes.  None of the existing experiments are sufficiently sensitive. For reference, strongly interacting massive particle (``SIMP''  \cite{Wandelt2001}) models are indicated by the black band.}
  \label{fig:sigma}
\end{figure*} 

\section{Dark matter search potential}

The search potential for low-mass \dm{} through decoherence is depicted in \Figref{fig:sigma}.  I will concentrate on the wide mass range 10 eV--100 MeV.  Above this range, conventional direct detection techniques will be superior.  Below this range, the occupation number of \dm{} momentum modes in the Milky Way surpasses unity. (Fermionic \dm{} would show signs of degeneracy, while bosonic \dm{} would behave like a coherent wave.)

The only existing direct-detection bound on the spin-independent nucleon-scattering cross-section for $\mdm < 1$ GeV comes from the X-ray Quantum Calorimetry experiment \cite{McCammonetal2002}, as analyzed by Erickcek et al.\ \cite{Erickcek2007}.  There is also a constraint arising from the stability of the \dm{} halo which encompasses the Milky Way, and the rate at which it heats interstellar hydrogen through collisions \cite{Chivukula1990}.  These robust exclusions are based only on the present-day distribution of the \dm{} which is necessary to explain observed galactic dynamics.

If one further assumes the simplest thermal freeze-out scenario for \dm{} in the early universe, tighter upper bounds on $\sigma$ for $\mdm > 1$ MeV have been derived from the cosmic microwave background and large-scale structure data \cite{Chen2002}. (Note that this limit can probably be extended to lower masses if more extensive analysis were done on the data.)  This type of thermal relic \dm{} also becomes too warm to explain small-scale structure data encoded in the Lyman-$\alpha$ forest when the mass falls below a few keV \cite{Viel2008}.  Both of these restrictions might easily be violated (by dark-observable temperature ratios \cite{Feng2008b} or fully non-thermal scenarios), as might other published bounds which rest on significant new assumptions about the nature of \dm{} (e.g.\ its self-annihilation or high-energy inelastic scattering to gamma rays \cite{Mack2012}).  

Note that it is possible for particle accelerators to probe the low masses discussed here (as well as the traditional masses sought by conventional direct detection experiments), but only when restricted to certain models.   For instance, by assuming a particular mediator and coupling one may look for trackless dijets \cite{bai2011dark} or monojets plus missing transverse energy \cite{rajaraman2011lhc, chatrchyanetal2012search,fox2012missing} at the LHC.

A quantum superposition experiment on Earth will not be sensitive to \dm{} if the scattering cross-section with nucleons is so large that the atmosphere shields the experiment from the \dm{} flux.  As shown in Fig.\ \ref{fig:sigma}, the maximum spin-independent cross-section visible to experiments on the Earth's surface is about $10^{-28.5} \mathrm{cm}^2$.  (This assumes that a single scattering event completely stops the \dm{}; if scattering is largely in the forward direction, the attenuation of the \dm{} flux might be much less.) To test \dm{} scenarios with larger cross-sections, the experiment could be operated on a high-altitude balloon ($\sim \!\! 30$ km altitude; $\sim \!\! 10^{-26.5} \mathrm{cm}^2$), a sub-orbital sounding rocket ($\sim \!\! 200$ km altitude; $\sim\!\! 10^{-20.5} \mathrm{cm}^2$), or a satellite.  The weight required to shield \dm{} in the range tested by \mbox{balloon-}, \mbox{rocket-}, and space-borne experiments is manageable. (Balloon: $10^{-28.5}\mathrm{cm}^2 \lesssim \sigma \lesssim 10^{-26.5}\mathrm{cm}^2$, $\ell_\mathrm{Pb} \lesssim 30$ cm.  Rocket and satellite: $10^{-26.5}\mathrm{cm}^2 \lesssim \sigma$, $\ell_\mathrm{Pb} \lesssim 3$ mm.)

I do not know if anyone has studied the possibility of producing large quantum superpositions on a balloon, but experiments on satellites and sub-orbital rockets are feasible and compelling for independent reasons. A microgravity platform offers several advantages for producing superpositions, whether using optical traps \cite{Kaltenbaek2012, kaltenbaek2013testing} or interferometers \cite{jentsch2004hyper, amelino-cameliaetal2008gauge,  ertmeretal2009matter, vanetal2010bose, Nimmrichter2011a, arndt2013free-falling, muntingaetal2013interferometry}. In particular, the unlimited free-fall times and isolation from seismic vibration available in orbit are able to increase sensitivity by multiple orders of magnitude and achieve quantum superpositions not feasible on Earth \cite{Kaltenbaek2012,sorrentinoetal2010compact, amelino-cameliaetal2008gauge, jentsch2004hyper, ertmeretal2009matter}.

Figure \ref{fig:sigma} shows the potential reach of several existing matter interferometers \cite{Carnal1991, Zeilinger1988, Brezger2002,Gerlich2011} in the absence of atmospheric shielding. The separation vector $\vec{\Delta x}$ is assumed to point into the \dm{} wind.  The effects of rotating $\vec{\Delta x}$ with respect to the wind are order unity and are depicted in \Figref{fig:chi}.  Modern experiments often use multiple gratings with many slits to overcome difficulties with beam coherence and tiny de Broglie wavelengths \cite{Cronin2009, Hornberger2011}, so the matter is not described by a simple superposition of two spatially separated wavepackets.  But the interferometers still require good coherence over distances which span multiple slits, so it is reasonable to estimate their sensitivity by taking $\Delta x$ to be the period of the relevant grating.  (Only the results for small $\mdm$ will depend on the choice of $\Delta x$; for larger masses, which are in the short wavelength limit, any scattering event results in complete decoherence independent of the spatial separation.)

To demonstrate the potential of future experiments to detect \dm{} through decoherence, I consider three proposals currently being developed.  First is the optical time-domain ionizing matter-wave (OTIMA) interferometer proposed by Nimmrichter et al. \cite{Nimmrichter2011a}.  An OTIMA interferometer eschews conventional material gratings for ionizing laser pulses. This avoids van der Waals interactions with the grating, and allows the superposition of potentially very large targets.  The design has recently been demonstrated with anthracene clusters larger than 2,000 amu, and will eventually interfere clusters of atoms that exceed $10^6$ amu \cite{Haslinger2013}.

Second is the interference of an optically-trapped dielectric sphere tens of nanometers in diameter proposed by Romero-Isart et al.\ \cite{Romero-Isart2011}.  In this case, the sphere would be laser cooled to its motional ground state and then dropped.  A laser pulse as it passed through a second cavity would prepare a spatial superposition which, after an additional fall, would then be confirmed through a position measurement.  (See Ref. \cite{kiesel2013cavity} for related progress.)

Third is the satellite-based Atomic Gravitational wave Interferometric Sensor (AGIS) proposed by Dimopoulos et al. \cite{hogan2011atomic}.  (See also Refs.\ \cite{Dimopoulos2008, dimopoulos2009gravitational, hohensee2011sources}.)   A pair (or triplet) of satellites would measure gravitational waves by operating widely separated atom interferometers with a common laser in low-Earth orbit.  The most likely configuration would involve rubidium-87 atoms superposed over a highly macroscopic distance of tens of meters for almost half a minute.

To get an intuition for the sensitivity of these experiments, one can estimate the lowest detectable cross-section in the large-$\mdm$ (short-wavelength) limit as
\begin{align}\begin{split}
\label{decoh-rate-est}
\sigma_0 \sim \frac{\mdm}{v_0 T \rho N A}
\end{split}\end{align}
where $A$ is the atomic mass number of the target nuclei and $N$ is the total number of nucleons in the target. 
This estimate can be obtained for a lone nucleon, $N = A = 1$, by taking the distance $v_o T$ traveled by a typical dark matter particle during the lifetime of the superposition and equating it to the mean free path $\mdm/\sigma \rho$ for a nucleon in the \dm{} ``gas''. 
The factor of $N$ accounts for the fact that \dm{} may decohere the superposition by striking \emph{any} of the $N$ nucleons in the target, and the factor of $A$ accounts for the scattering enhancement (discussed in detail in the Appendix) due to coherence across individual nuclei.  We have $\abs{\gamma}  \sim 1/e$ when $\sigma = \sigma_0$.

In the opposing low-mass limit, where $\lambda_0 \gg \Delta x$, the detectable threshold \eqref{decoh-rate-est} is raised by a factor $(\Delta x/\lambda_0)^2$ due to the indistinguishability effects illustrated in Fig.\ \ref{fig:wave}.  Additionally, $A$ is replaced by a second factor of $N$ to account for coherent scattering across the entire target object.

The AGIS satellite would interfere atoms in the open vacuum of space, so it would be sensitive to \dm{} scenarios for $\mdm \lesssim 3$ keV which have never been excluded, as shown in \Figref{fig:sigma}. The OTIMA and nanosphere experiments would need to be raised at least partially out of the Earth's atmosphere to see dark matter, but experiments further into the future could rule out \dm{} scenarios at ground level. For $\mdm \lesssim 3$ keV, there is a significant region for which \dm{}'s unitary phase shift can be observed in the OTIMA and nanosphere experiments without being overwhelmed by decoherence.

It is likely \cite{Kuhlen2010} that the true \dm{} velocity distribution has a thicker tail than the Maxwellian form assumed here.  High momentum \dm{} causes disproportionately more decoherence, so this should increase sensitivity further. Also note that if there were a mechanism which increased the local \dm{} density in the vicinity of the Earth (e.g.\ Ref.\ \cite{Adler2009}), this would improve the sensitivity of decoherence detection and other near-Earth methods without changing the astrophysical limits which currently provide the best bounds.

Even more importantly, statistical analysis of decoherence rates over many runs of the experiment may be able to increase the sensitivity  by several orders of magnitude.  This could be particularly convincing if the cross-section is large enough ($\sigma \gtrsim 10^{-30} \mathrm{cm}^2$) to be shielded by moveable barriers.  In this case, well controlled trials could be performed to search for slight increases in the decoherence rate when shielding is removed.  For $M$ targets passed through the interferometer, the cross-section sensitivity scales like $\sqrt{M}$.  Since count rates for typical matter interferometers (which have \emph{not} been optimized for dark matter) are on the order of thousands per second, the potential increase in sensitivity from data collected over several months is significant.  Experiments that cannot be manually shielded would have to rely on the natural shielding of the Earth or on the directional sensitivity to rotation.  (These trials would not be as well controlled, and confounding factors correlated with the method of shielding might be introduced.)  The AGIS satellite is likely to individually measure $\sim 10^8$ atoms per shot \cite{hogan2011atomic} and so also has a large potential for enhancement through statistics.

\section{Discussion}

It's worth stressing that detection through decoherence is not limited to interferometers.  In principle, any superposition of states of normal matter separated in phase-space is sensitive to collisional decoherence from \dm{}. Larger objects composed of many particles are especially so.

Macroscopic superpositions of mechanical oscillators \cite{OConnell2010, Romero-Isart2012, Pepper2012} are promising because of the sheer size of the masses under quantum control.  These are very different than the traditional interferometers depicted in \Figref{fig:sigma} because energy (rather than position) eigenstates are superposed, and because the rough separation $\Delta x$ is much smaller than the oscillators themselves.  The latter fact means the spatial size of the individual nuclei targets must be considered and compared to $\Delta x$ since \dm{} scattering from locations common to the two eigenstates will not decohere. Rough estimates suggest that some of these proposed devices would have even larger sensitivity to \dm{}, although a detailed analysis has not been performed.

On the other hand, Bose-Einstein condensates (BECs) are not naturally suited for detection by decoherence.  Spatial interferometry has been done with BECs \cite{Schumm2005, ertmeretal2009matter, van2010bose}, but this is essentially \emph{atom} interferometry; the atoms in a BEC are all in the same state, but they aren't entangled.  Individual atoms in the BEC can be lost without destroying the coherence, so there is no boost in sensitivity like there is for superposing the center-of-mass of large clusters of atoms.  (For the same reason, normal BECs do not avoid the shot noise limit when used to measures phase differences.) However, the creation of entanglement in BECs, such as NOON \cite{Cable2011} or spin-squeezed \cite{MaxRiedel2010} states, might be exploitable.

The scattering cross-section of \dm{} with electrons, rather than nucleons, is also of interest \cite{Essig2012a, Graham2012}.  It could be probed with the matter interferometers discussed in this article, requiring additional analysis but no modification to the experiments. It might also be investigated with superconducting qubits, in which two experimentally manipulable quantum states are composed of millions of entangled Cooper pairs \cite{clarke2008superconducting}.  In the case of flux qubits, scattering \dm{} could record ``which-momentum'' information about the electrons in these macroscopic states, and so decohere them.

All of the experiments discussed in this article were performed or proposed for reasons completely independent of \dm{} detection.  It is likely that their \dm{} sensitivity can be significantly improved were they designed with that in mind \cite{ArndtPC}.

Beyond dark matter, one can reinterpret many experiments which establish certain quantum states as direct evidence against hypothetical weak phenomena that, if existent, would decohere those states.  The toy Mach-Zehnder interferometer illustrates that the classical effects of such phenomena (e.g.\ momentum transfer) can be arbitrarily small while still causing very noticeable decoherence.  The potentially extreme detection sensitivity of macroscopic superpositions gives new independent motivation for their experimental pursuit.

\begin{acknowledgments}

I thank Markus Arndt, Asimina Arvanitaki, Dirk Bouwmeester, Xiaoyong Chu, Savas Dimopoulos, Rouven Essig, Alexander Friedland, Jay Gambetta, Christian Hagmann, Steen Hannestad, Andrew Hime, Lorenzo Maccone, Gregory Mack, Jeremy Mardon, Benjamin Monreal, Kevin Moore, Harry Nelson, Bryon Neufeld, Shmuel Nussinov, Keith Rielage, Daniel Sank, Robert Scherrer, Alexia Schulz, Mark Srednicki, Paul Steinhardt, Lev Vaidman, Neal Weiner, Haibo Yu, and Yue Zhao for discussion and comments.  I am especially grateful to Michael Graesser for teaching, Godfrey Miller for criticism, and Charlie Bennett, Jim Hartle, and Wojciech Zurek for making this work possible. This research was partially supported by the U.S. Department of Energy through the LANL/LDRD program, and by the John Templeton Foundation through grant number 21484.

\end{acknowledgments}

\bibliographystyle{apsrev4-1}
\bibliography{riedelbibetal}

\appendix*

\section{Coherent scattering enhancement}

Here I discuss the enhanced scattering cross-section due to coherent elastic scattering for targets composed of many atoms where the dark matter (DM) de Broglie wavelength is comparable to or larger than the atomic spacing.  Coherent elastic scattering is known to play a crucial role in small-angle scattering experiments with neutrons and X-rays \cite{SquiresText}, and in the as-yet unobserved scattering of relic (or cosmic) neutrinos \cite{Smith1991}.  The case of DM is especially analogous to relic neutrinos, which are widely believed to have been produced within a few seconds of the Big Bang. Like the comic microwave background, relic neutrinos have freely streamed through the universe ever since they decoupled.  The neutrino background is thermal, and the expansion of the universe has stretched the de Broglie wavelength to truly macroscopic distances nearing the order of a millimeter.  The neutrino-nucleon cross-section depends on momentum---and so, because they are likely non-relativistic, on the neutrino masses---but is in any case extremely tiny ($<10^{-52}\, \mathrm{cm}^2$).  Although ultimately judged to be unfeasible for the foreseeable future, there were several theoretical investigations of the possibility of detecting relic neutrinos which relied on the tremendous cross-section enhancement when neutrinos scatter coherently from essentially macroscopic targets.  In particular, target granules larger than 0.01 mm in diameter have been analyzed and the scattering of relic neutrinos is expected to be fully coherent across the entire target \cite{Smith1991}.

The case of DM scattering from targets placed in a quantum superposition is similar, but differs crucially in that no momentum transfer is required.  Consider $N$ nucleons in an amorphous target of volume $V$ with $N_a = N/A$ identical atoms with mass number $A$ located at positions $\vec{x}_i$. If the different nucleons were to contribute to the decoherence rate 
\begin{align}\begin{split}
\label{decoh-rate-appendix}
F_\mathrm{R}(\vec{\Delta x}) =& \int \dd \vec{q} \, n(\vec{q}) \frac{q}{\mdm} \int \dd  \hat{r} \\
& \times \left\{1 - \cos[(\vec{q} -  q\hat{r}) \cdot \vec{\Delta x}/\hbar]\right\} \frac{\sigma}{4 \pi}.
\end{split}\end{align}
incoherently (such as if the \dm{} were to flip the spin of a nucleon), then $F_\mathrm{R}$ would simply be multiplied by a factor of $N$ compared to the case of a single nucleon.  But in the coherent case appropriate to the sub-MeV \dm{} discussed in this article, \eqref{decoh-rate-appendix} is modified within the Born approximation by inserting a structure factor \cite{SquiresText}
\begin{align}\begin{split}
\label{boost-factor}
I(\vec{\Delta q}) &= \Avg{\Abs{\sum_{i=1}^{N_a} A \, e^{-i \vec{x}_i \cdot \vec{\Delta q}/\hbar}}^2} \\
&= A^2 \Avg{\sum_{i=1}^{N_a} \sum_{j=1}^{N_a} e^{-i (\vec{x}_i-\vec{x}_j) \cdot \vec{\Delta q}/\hbar}}
\end{split}\end{align}
inside both integrals.  Here, $\vec{\Delta q} = \vec{q} - q \hat{r}$ is the momentum transfer and $\avg{\cdot}$ denotes a thermal average. This structure factor is only sensitive to the distribution of atoms in the target through the one- and two-particle probability distributions. Neglecting edge effects, $p(\vec{x}_i,\vec{x}_j) = p(x_i) p(x_j) g(\abs{\vec{x}_i - \vec{x}_j})$ where $g(r)$ is the pair-correlation function.

\begin{figure*} [tb!]
  \centering 
\newcommand{\donfactor}{1.01}
\includegraphics[width=\donfactor\columnwidth]{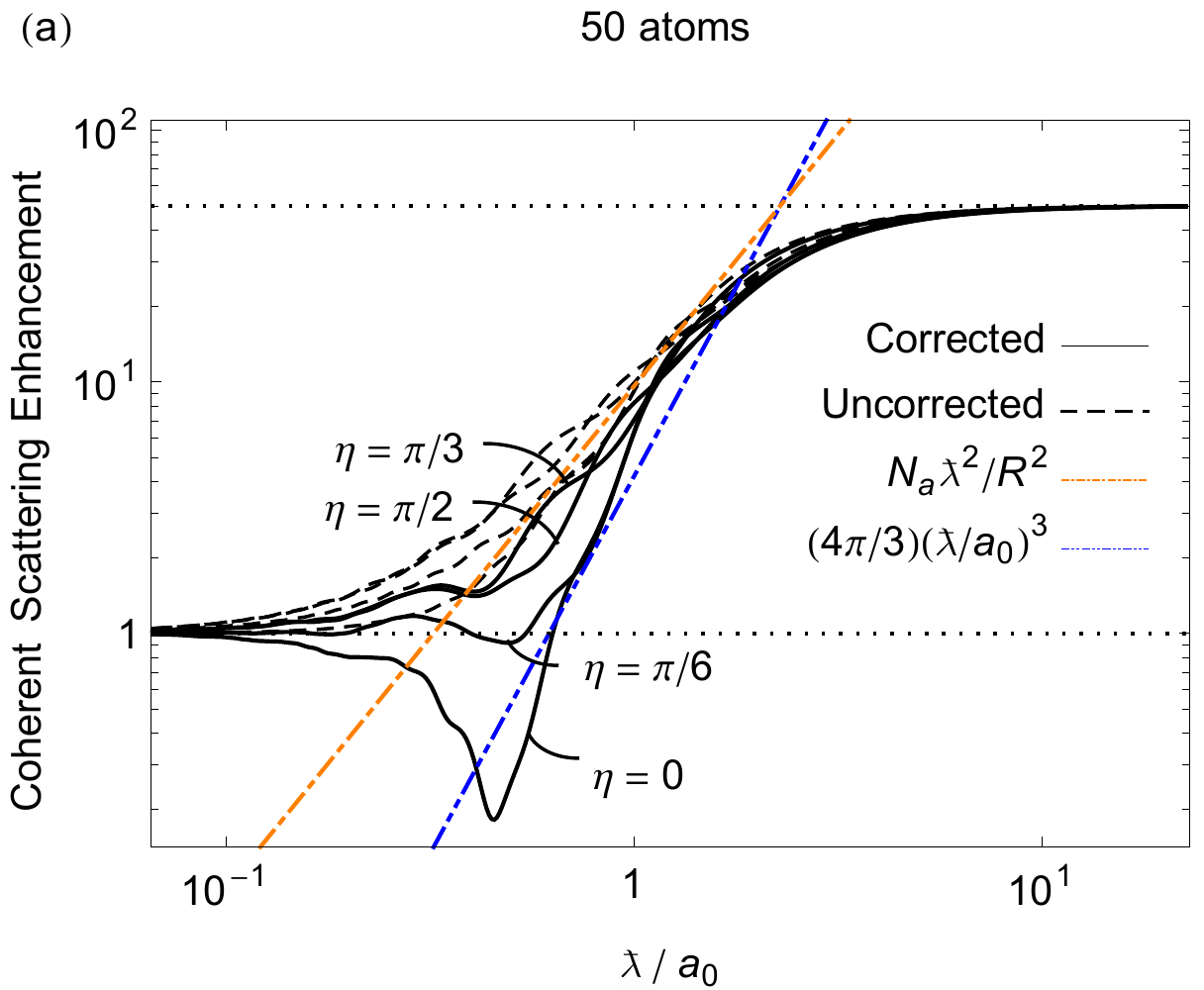}
\includegraphics[width=\donfactor\columnwidth]{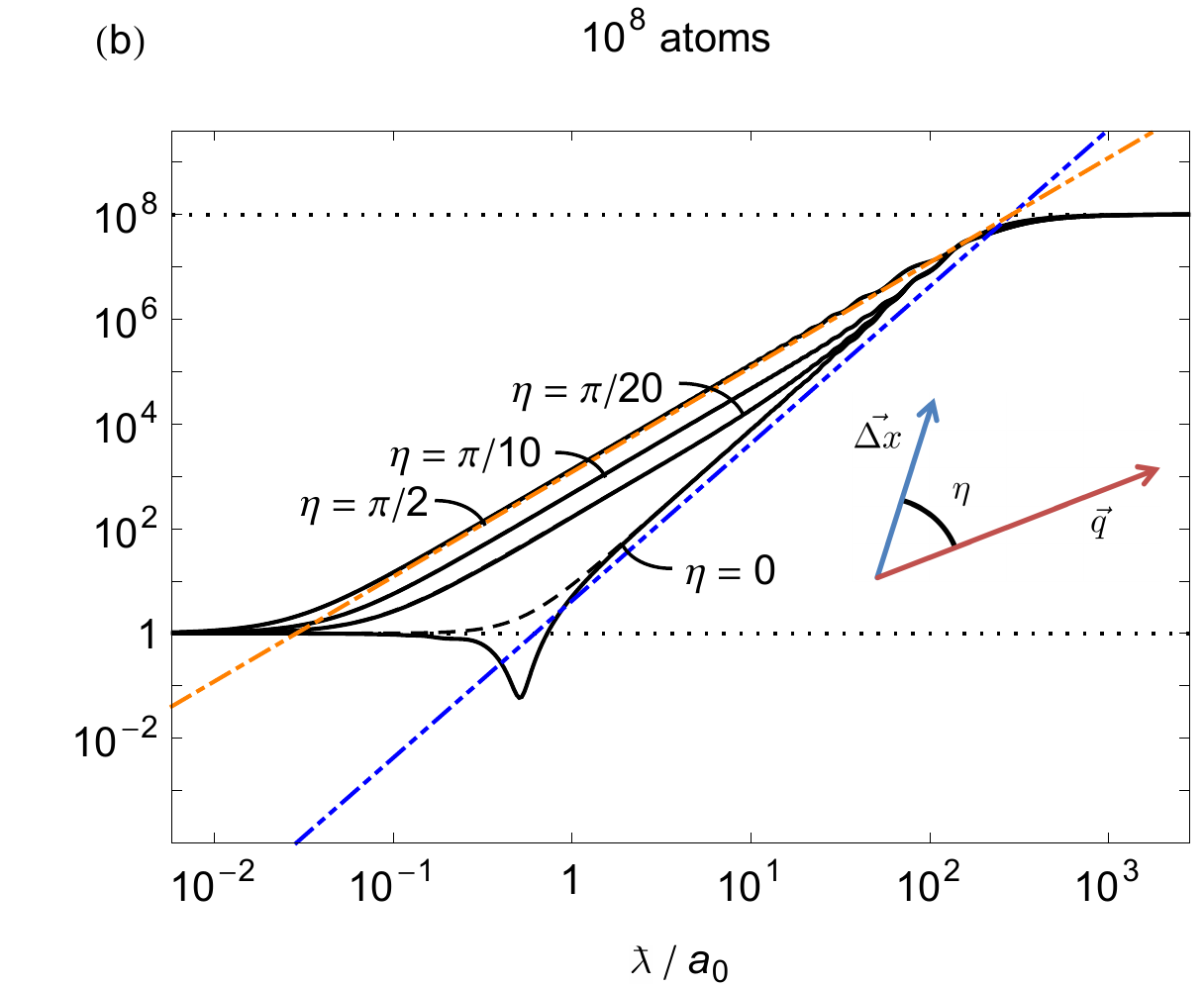}
\caption{\textbf{Coherent scattering enhancement.} (a) The coherent scattering enhancement $B_\mathrm{coher} (\vec{q})/A^2$ to a single incoming \dm{} particle's contribution to the decoherence rate $F_\mathrm{R}$ relative to incoherent scattering.  (The effects of the fixed mass number $A$ have been factored out.) The magnitude of the incident momentum $\vec{q}$ is given by $q = \hbar/\lambdabar$ and the polar angle $\eta$ is measured with respect to the superposition separation $\vec{\Delta x}$.  The target is taken to be $N_a = 50$ atoms in a spherical cluster of radius $R = [3V/4\pi]^{1/3}$, which is superposed over a distance $\Delta x$ equal to the diameter.  The reduced wavelength $\lambdabar$ is measured in units of the atomic spacing $a_0$.  Four values of $\eta$ are considered.  ($B_\mathrm{coher}$ is invariant under $\eta \to \pi - \eta$ because $\vec{\Delta x} \to - \vec{\Delta x}$ is physically equivalent.)  The solid (dashed) black lines depict the enhancement calculated with (without) corrections from the pair-correlation function $g(r)$.  The blue dash-dotted line gives the average number of atoms which fit in the coherent scattering volume $(4 \pi/3) \lambdabar^3$. It provides a good (and almost always conservative) estimate of the coherent scattering enhancement.  The orange dash-dotted line denotes $N_a (\lambdabar^2/R^2)$, the form of which can be derived analytically from \eqref{eq:boost} in the $R \to \infty$ limit. The horizontal dotted lines denote full coherence and incoherence ($B_\mathrm{coher} = N_a,1$). (b) The  enhancement for a spherical cluster of $N_a = 10^8$ atoms.  The corrections from the pair correlation function are small because $a_0 \ll R$.  As target size increases, the $B_\mathrm{coher}$ curve is pushed toward the orange $\lambdabar^2$ line except when $\vec{\Delta x}$ and $\vec{q}$ are nearly aligned.  This behavior can be traced to the oscillatory nature of the term in curly braces in \eqref{eq:boost}.}
  \label{fig:boost}
\end{figure*} 

If the correlations are trivial, $g(r) = 1$, then each position $\vec{x}_i$ is uniformly and independently distributed over the volume.   For a target with characteristic spatial size $L$ which is much smaller than $\lambdabar = \hbar/q$, the dot product in the exponent is always much less than unity so $I(\vec{\Delta q})$ evaluates to $N^2 = N_a^2 A^2$, i.e.\ an enhancement of $N$ compared to the incoherent case. On the other hand, if $\lambdabar$ is sufficiently small compared to the atomic spacing $a_0=(V/N)^{1/3}$, then each term on the right-hand side of \Eqref{boost-factor} will vanish under the thermal average except when $i=j$, yielding $I(\vec{\Delta q}) = N_a A^2$.  (There is always an enhancement of $A$ compared to the fully incoherent case due to coherence across the individual nuclei.  This only breaks down when $\lambdabar$ approaches the nuclear scale, which is not a concern for decoherence-based \dm{} detection schemes since they are only useful when $\mdm < 1$ GeV.)

In the intermediate case---when the wavelength is large enough to span multiple nuclei but not large enough to span the whole target---we can numerically calculate the effective boost to the decoherence rate by a single incoming \dm{} particle by integrating the structure factor $I(\vec{\Delta q})$ and the overlap term in curly braces in \eqref{decoh-rate-appendix} over all possible out-going directions $\hat{r}$ and normalizing by the same without the structure factor:
\begin{align}
\label{eq:boost}
B_\mathrm{coher} (\vec{q}) = \frac{\int \dd  \hat{r} I(\vec{\Delta q}) \left\{1 - \cos[\vec{\Delta q} \cdot \vec{\Delta x}/\hbar]\right\} }{\int \dd  \hat{r} \left\{1 - \cos[\vec{\Delta q} \cdot \vec{\Delta x}/\hbar]\right\} }.
\end{align}
This is plotted in Fig.\ \ref{fig:boost} where, for simplicity, $g(r)$ is taken to vanish inside a radius of $a_0$ and is constant outside, reflecting the fact that atoms cannot lie on top of each other but are otherwise nearly uncorrelated in an amorphous solid.  This illustrates that approximating the enhancement as the number of nucleons which fit in the coherent scattering volume $\lambdabar^3$ is a reasonable practice when estimating the DM sensitivities of interferometers, especially as target size increases.  The approximation is conservative in the sense that the actual sensitivity is generally greater.  (The only exception is the suppression that happens when the incoming DM momentum $\vec{q}$ is very closely aligned with the superposition separation $\vec{\Delta x}$, but this effect will be swamped by contributions from the rest of the \dm{} flux.)

For large targets ($L \gg \lambdabar \gtrsim a_0$), most of the scattering is in the forward direction due to destructive interference in the structure factor when $\Delta q /q \gtrsim \lambdabar/L$. (Note that the Born approximation applies for arbitrarily small momentum transfers \cite{Zemach1958, TaylorTextExcerpt}.) Although traditional small-angle scattering with neutrons and x-rays is limited by detector acceptance and angular spread of the incident beam, this is not a concern for causing decoherence.  Instead, the separation $\Delta x$ of the superposition (which is taken in this article to always be larger than the size $L$ of the target) limits decoherence through arbitrarily small-angle scattering by way of the term in curly braces in \eqref{eq:boost}. 

On the other hand, the effectiveness of \dm{} shielding is determined by the degree to which incident \dm{} momentum is attenuated, not by the decoherence it effects.  There is still an enhancement of $A$ due to coherence across the nucleus, but the relative phases between nuclei within a bulk material are not correlated (except for scattering in the forward, non-attenuating direction).  Their contributions to shielding therefore add incoherently.

(This also explains why target granules are not chosen to be larger than the neutrino de Broglie wavelength for the purposes of relic neutrino detection.  The signal of neutrino scattering would be spatial displacement of the granules, so momentum transfer is necessary.  As described above, momentum transfer quickly vanishes as the granule size surpasses $\lambdabar$ because outgoing neutrino states would interfere destructively in all directions besides the forward direction.)

It's illuminating to see this bulk limit $L/\lambdabar \to \infty$ explicitly.   For simplicity take the pair-correlation function to be trivial and, like for Fig.\ \ref{fig:boost}, assume the target is a sphere of radius $R = [3 V /4 \pi]^{1/3}$. Then
\begin{align}\begin{split}
I(\vec{\Delta q}) &= A^2\left[\sum_{i=1}^{N_a} \Avg{1} + \sum_{i=1}^{N_a} \sum_{\substack{j=1\\j \neq i}}^{N_a} \Avg{e^{-i (\vec{x}_i-\vec{x}_j) \cdot \vec{\Delta q}/\hbar} }\right] \\
&= A^2\left[N_a + \frac{N_a-1}{N_a} \rho^2  \left| \int \ddd{\vec{x}}  e^{-i \vec{x} \cdot \vec{\Delta q}/\hbar} \right|^2 \right] \\
&= A^2\bigg[N_a + \frac{N_a-1}{N_a} \left(4 \pi \rho f(R \Delta q/\hbar) R^3 \right)^2 \bigg].
\end{split}\end{align}
where the integral in nuclei positions $\vec{x}$ is taken over the target volume and where
\begin{align}
f(s) \equiv \frac{\sin(s)-s \cos(s)}{s^3}.
\end{align}
The cross-section for all events with scattering angle $\theta \ge \bar{\theta}$ is 
\begin{align}\begin{split}
\sigma_{\mathrm{bulk}}^{\ge \bar{\theta}} &= \int_{\theta \ge \bar{\theta}}  \frac{\dd  \hat{r}}{4 \pi} \sigma I(\vec{q}-q\hat{r}) \\
&= \sigma A^2 \left[ N_a +  N_a(N_a-1) \frac{9}{4 \pi}  \int_{\theta \ge \bar{\theta}} \dd  \hat{r} \, f(R \Delta q/\hbar)^2 \right] .
\end{split}\end{align}
For fixed $\lambdabar = \hbar/q$ and sufficiently large $R$, one can show that the integral over outgoing directions $\hat{r}$ falls like $1/R^2$ when $\bar{\theta}$ vanishes exactly but like $1/R^4$ for any $\bar{\theta} > 0$.  This is because the scattering gets more and more focused in the forward direction as $R$ increases. Since shielding is only effective insofar as it attenuates the initial momentum, only strictly positive values of $\bar{\theta}$ are relevant.  The second term inside the square brackets then becomes negligible for large $R$ (since $N = \rho V = 4 \pi \rho  R^3/3$), and we recover $\sigma_{\mathrm{shield}} = N_a A^2  \sigma$.  In other words, the only enhancement relevant to shielding by bulk materials is given by the mass number $A$ of the nuclei.

For fixed $R$, on the other hand, 
\begin{align}
\lim_{\lambdabar \to \infty} \frac{9}{4 \pi}  \int_{\theta \ge \bar{\theta}} \dd  \hat{r} \, f(R \Delta q/\hbar)^2 = \frac{1+\cos \bar{\theta}}{2}, 
\end{align}
which is unity for $\bar{\theta} = 0$.  When $\vec{\Delta x}$ is sufficiently large (compared to the fixed $R$) that the cosine term in \eqref{decoh-rate-appendix} averages to zero, then the decoherence rate is directly proportional to the total cross-section ($\bar{\theta} = 0$) of the target.  We recover $\sigma_{\text{target}} = N_a^2 A^2  \sigma = N^2 \sigma$.  That is, a sufficiently large coherent scattering volume $\lambdabar^3$ guarantees a full $N^2$ enhancement (as expected).

Finally, one can check the Debye-Waller factor to confirm that treating the nuclei as rigidly fixed scattering centers is appropriate \cite{SquiresText}.  For a given momentum transfer, this factor is given by $\exp(-\Delta q^2 \Avg{u^2}/3\hbar)$ where $\Avg{u^2}$ is the thermal mean-squared displacement of the nuclei motion in the target at the appropriate temperature.  The Debye-Waller factor quantifies the degree of suppression of coherent elastic scattering due to inelastic interactions with phonon modes in the target. The mean-squared displacement can be approximated using the Debye model and, for temperatures $T$ above roughly 100 K, one gets 
\begin{align}
\Avg{u^2} \approx \frac{4 k_\mathrm{B} T }{\pi c_\mathrm{s}^2  a_0 \rho } = d_0^2 \left( \frac{T}{300 K} \right)
\end{align}
where $c_s$ is the speed of sound in the target, and $\rho$ is the target mass density.  (For low temperatures, $\Avg{u^2}$ approaches a positive minimum value set by the zero-point motion of the atoms.) For gold, this evaluate to $d_0 \approx 0.1$~\AA{} while the atomic spacing is $a_0 \approx 2.6$~\AA{}. Since we are only interested in \dm{} wavelengths $\lambdabar$ of order the atomic spacing or greater, the Debye-Waller factor $\exp(-\Delta q^2 \Avg{u^2}/3\hbar)$ is close to unity for relevant temperatures of the multi-atom target experiments depicted in Fig.\ \ref{fig:sigma}.  This means the coherent elastic scattering cross-section is not substantially suppressed due to the inelastic excitement of phonon modes in the target.

\end{document}